\documentclass[fleqn,10pt]{wlscirep}
\usepackage{geometry}
\usepackage{etoolbox}
\usepackage{graphicx}
\usepackage{color}
\usepackage{graphicx}
\usepackage{amsmath}
\usepackage{capt-of}
\usepackage{caption}
\usepackage{slashbox}
\usepackage{bm}

\title{Enhanced and stable spin Hall conductivity in a disordered time-reversal and inversion symmetry broken topological insulator thin film}

\author[1]{Siamak Pooyan}
\author[1,*]{Mir Vahid Hosseini}
\affil[1]{Department of Physics, Faculty of Science, University of Zanjan, Zanjan 45371-38791, Iran}
\affil[*]{mv.hosseini@znu.ac.ir}

\begin{abstract}
We consider a disordered topological insulator thin film placed on the top of a ferromagnetic insulator with a perpendicular exchange field $M$ and subjected to a perpendicular electric field. The presence of ferromagnetic insulator causes that bottom surface states of the topological insulator thin film become spin polarized and the electric field provides a potential difference $V$ between the two surface states, resulting in breaking of time-reversal and inversion symmetry in the system. Using Kubo formalism and employing the first Born approximation as well as the self-consistent Born approximation, we calculate the spin Hall conductivity. We find that for small values of $V$, a large spin conductivity can be generated through large values of $M$ away from the charge neutrality point. But for large values of $V$, the spin conductivity can be promoted even with small values of $M$ around the charge neutrality point. The effect of vertex corrections and the stability of the obtained large spin conductivity against disorders are also examined.
\end{abstract}

\begin{document}
\flushbottom
\maketitle

\thispagestyle{fancy}

%%%%%%%%%%%%%%%%%%%%%%%%%%%%%%%%%%%%%%%%%%%%%%%%%%%%%%%%%%%%%%%%%%%%%%%%%%%
\section {Introduction} \label{s1}
Topological insulators (TIs) have attracted a lot of attentions from theoretical viewpoint and potential applications \cite{Topo1,Topo2,Topo3,Topo4,Topo5,Topo6}. An interesting feature of these matters is that, in a topologically nontrivial phase, edge or surface states of the system exhibit metallic feature, due to intersecting the Fermi level, while bulk states are an ordinary insulator at the Fermi level \cite{Topobound1,Topobound2}. Nontrivial topology of bulk states can be related to the appearance of surface states, resulting in the bulk-edge correspondence. The topology can be supported by certain symmetries of the system. This causes surface states become robust against perturbations respecting essential symmetries of the system manifesting symmetry-protected topological states.

When the thickness of a 3D TI decreases so that it becomes five to ten quintuple layers \cite{thickness1,thickness2}, topological states of the opposite surfaces can hybridize together providing an interesting opportunity for applications. Because of the hybridization, a gap opens in the surface spectrum and carriers behave like massive Dirac fermions in a thin film of TIs \cite{thickness3,thickness4}. This feature is expected to lead to the spin Hall effect in TI thin films \cite{SHEinThinFilm}.

The spin Hall effect \cite{spinHall} can occur due to a large spin-orbit coupling in time-reversal symmetric TIs. As such, a high spin conductivity would be expected in TIs owing to the prohibition of back scatterings. TIs have also been studied in a magnetic field manifesting the magnetoelectric coupling \cite{magnetoelectric} and the magnetoresistance \cite{magnetoresistance}. Moreover, in a strong magnetic field perpendicular to the surface of a TI, the charge Hall effect can coexist with the spin Hall effect \cite{charge-spin-Hall}. Recently, the study of magnetic properties of TI thin film has become one of the hot topics in this issue. It has been shown that TI thin films can exhibit the giant magneto-optical Kerr effect \cite{GiantMagneto}, the topological magnetoelectric effect \cite{TopoMagEle}, and the giant magnetoresistance \cite{giantMagResi} with large spin Hall angles \cite{magnetoTranLarge}.

On the other hand, disorders, including impurities and defects, are ubiquitous, particularly, in solid state materials, e.g., TIs. A competition between charged impurity scattering and short-range scattering in TIs with hexagonal warping \cite{bulk} has been studied \cite{ImpScattTI}. In the presence of nonmagnetic and magnetic disorders, respectively, a large out-of-plane and an in-plane magnetoresistance have been found on the surface of 3D TIs that is proximity-coupled to a ferromagnetic insulator (FI) \cite{SpinTransportDisorFI-TI}.
In the thin TI film, surface electrons can screen a disorder potential that is larger than the hybridization gap \cite{ImpSceenThinTIFilm}. However, the spin Hall effect is a fragile state and its experimental measurement is a challenging task due to ubiquitous impurities and imperfections in samples \cite{spinHall}. So, it is interesting to promote the spin Hall conductivity with a stable character against disorders paving the way in spintronic applications.

In this respect, while in most of previous cases a magnetic exchange field is applied to the whole system, including both surfaces of the TI thin film, it is interesting to know what happens if the magnetization affects on only one of the surfaces of thin film. In the present work, we calculate the dc spin conductivity in a one-surface-magnetized TI thin film by the Kubo formalism with random nonmagnetic potential disorders. To induce the magnetization in a one of the surfaces of TI thin film, one can attach a ferromagnet to a surface of the thin film. This also can provide a potential difference between the top and bottom surfaces. We use the first Born approximation (fBA) and the self-consistent Born approximation (SCBA) to treat nonmagnetic disorders, respectively, analytically and numerically. We calculate the self-energies, density of states (DOS), spin conductivity, and vertex-corrected velocity function. Interestingly, we find that the spin Hall conductivity can be enhanced for a large (small) exchange field and small (large) potential difference in a large (small) chemical potential. Also, it is shown that the promoted spin Hall conductivity remains survived at high enough impurity strengths that is a key requirement in spintronics.

The rest of the paper is organized as follows. In Sec. \ref{s2}, we present model and theory for the surface states of TI thin films with a magnetic exchange field in a one surface and a potential difference between the two surfaces. In Sec. \ref{s3}, disorders are modeled by nonmagnetic impurities on the two surfaces and included in self-energies using the fBA and the SCBA. The effect of disorders on DOS is also studied. In Sec. \ref{s4}, the Kubo formula is used to obtain spin Hall conductivity in the system. Section \ref{s5} is devoted to studying the effect of vertex corrections. We summarize in Sec. \ref{s6}.

%%%%%%%%%%%%%%%%%%%%%%%%%%%%%%%%%%%%%%%%%%%%%%%%%%%%%%%%%%%%%%%%%%%%%%%%%%%
\section {Model}\label{s2}
%%%%%%%%%%%%%%%%%%%%%%%%%%%%%%%%%%%%%%%%%%%%%%%%%%%%%%%%%%%%%%%%%%%%%%%%%%%
\begin{figure}[t]
\centering
\includegraphics[width=8cm]{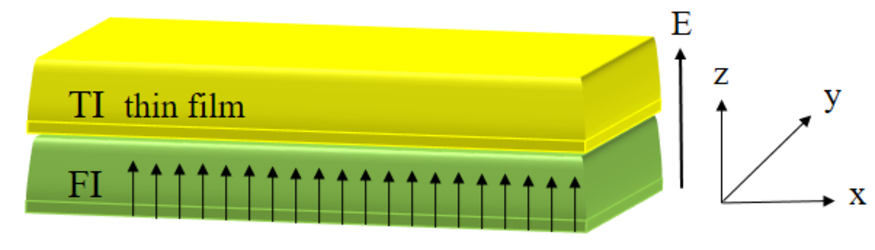}
\caption{(Color online) Schematic of a TI thin film placed on the top of a FI in the presence of an electric field. The surfaces of the film are on the $xy$-plane and both the polarization of FI and the electric field are along the $z$ direction.}
\label{thinfilm}
\end{figure}

We consider a TI thin film, having two surface states localized at the opposite surfaces, placed on the top of a FI in the presence of an electric field, see Fig. \ref{thinfilm}. The low-energy effective Hamiltonian of the system can be written as \cite{EffecHamil1,EffecHamil2,EffecHamil3}
\begin{equation}
H=rk^2+(H_R-V)\tau_z+\Delta \tau_x-M\sigma_z\tau_-,
\label{hamil}
\end{equation}
with
\begin{equation}
H_R = v_{Fk}(k_{x} \sigma_{y}- k_{y}\sigma_{x}),
\label{Rash}
\end{equation}
where $v_{Fk}= v_{F}(1+sk^2)$ with $v_{F}$ being Fermi velocity and $s$ characterizes the next order correction to the Fermi velocity. $k=\sqrt{k_x^2+k_y^2}$ with $k_{x}=k \cos \phi$ and $k_{y}=k \sin \phi$ being the in-plane momentum components. The Pauli matrices $\boldsymbol \sigma$ and $\boldsymbol \tau$ act on the spin and the surface spaces, respectively. $\tau_{\pm}=(\tau_0{\pm}\tau_z)/2$ with $\tau_0$ being a unit matrix. $r=1 /(2 m)$ is the inverse mass term originates from asymmetry between the electron and hole bands \cite{EffecHamil1}. Also, $V$ is the potential difference between the top and bottom surfaces, and $\Delta$ is the hybridization matrix element between top and bottom surface states \cite{Hybrid}. $M$ is the exchange field due to FI being applied to the bottom surface of TI. In this system, the inversion and time-reversal symmetry are broken by the electric and exchange fields. These fields are assumed to be perpendicular to the surface of TI film, i.e., along the normal direction $z$. The effect of an in-plane exchange field \cite{SpinTransportDisorFI-TI} which shifts the position of the Dirac points oppositely has been studied in the TI thin film leading to the giant anisotropic magnetoresistance at low dopings \cite{AnisoMagneResTITF}.
\begin{figure}[t]
\centering
\includegraphics[width=8cm]{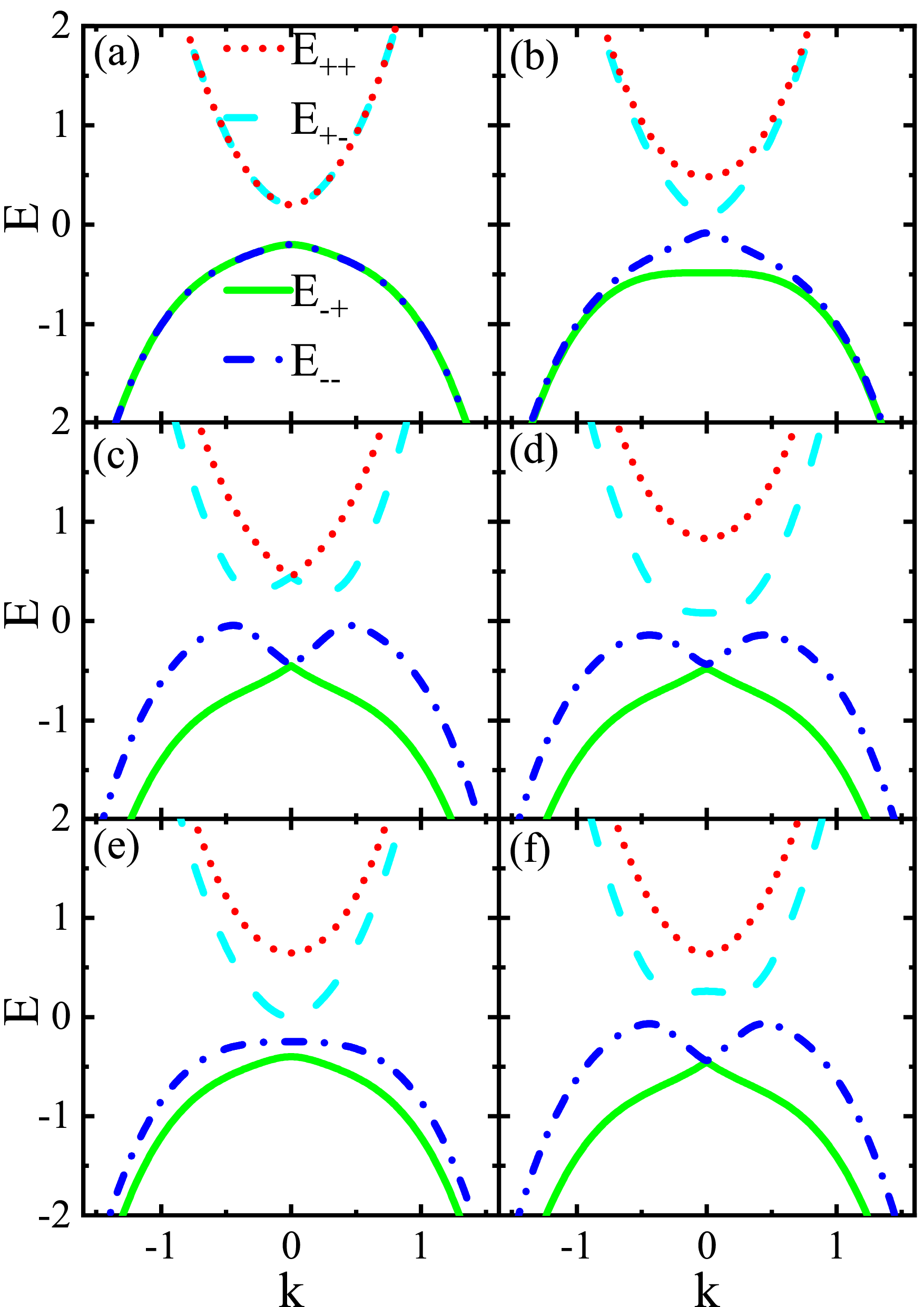}
\caption{(Color online) Energy spectra as a function of the $k$ for (a) $M=0$ and $V=0$, (b) $M=0.4$ and $V=0$, (c) $M=0$ and $V=0.4$, (d) $M=0.4$ and $V=0.4$, (e) $M > V$ with $M=0.4$ and $V=0.2$, and (f) $M < V$ with $M=0.2$ and $V=0.4$. Here $\Delta = 0.2$, $s=1$, and $r=1$. }
\label{fig2}
\end{figure}

Although Hamiltonian (\ref{hamil}) cannot be diagonalized analytically, one can obtain the energy spectra numerically. In special cases, however, the energy spectra can be obtained as
\begin{equation}
E_{lp}=r k^{2}+ l \sqrt{V^{2}+v_{Fk}^{2} k^{2}+\Delta^{2}+ 2 p v_{Fk} k V},
\label{eig1}
\end{equation}
for $M=0$ and $V\neq 0$ and
\begin{equation}
E_{lp}=r k^{2} +l \sqrt{\frac{M^{2}}{2}+v_{Fk}^{2} k^{2}+\Delta^{2}+ p \frac{\sqrt{M^{4}+4 M^{2} \Delta^{2}}}{2}},
\label{eig2}
\end{equation}
for $M\neq0$ and $V=0$. Here, $l=\pm$ stand for conduction and valence bands and $p=\pm$ indicate different subbands. Throughout the paper, we take $v_{F}/a$ as the unit of energy with $a$ being the lattice constant as the unit of length.

The energy spectra of Hamiltonian (\ref{hamil}) as a function of the $k$ are depicted in Fig. \ref{fig2} for different values of the $M$ and $V$. For $M = V = 0$, the band structure is gapped and the doubly degenerate conduction and valence bands, respectively, have a minimum and a maximum at $k=0$, see Fig. \ref{fig2}(a). As shown in \ref{fig2}(b), for $M \neq 0$ and $V = 0$, the band degeneracy is lifted for the states near $k=0$. Also, interestingly, the gap between the conduction and valence band decreases and tends to zero when $M \gg \Delta$. This is in contrast to the case where the exchange field $M$ is applied to both surfaces of TI thin films \cite{EffecHamil2}. On the contrary, for $M = 0$ and $V \neq 0$, the double degeneracy of bands breaks except at state $k=0$, see Fig. \ref{fig2}(c). Furthermore, the minimum (maximum) of lower conduction (upper valence) band splits into two minimums (maximums) located at finite $k$.
For finite values of both the $M$ and $V$, the band structure is shown in Fig. \ref{fig2}(d), Fig. \ref{fig2}(e), and Fig. \ref{fig2}(f), respectively, when $M = V$, $M > V$, and $M < V$. When $M \leq V$, the degeneracy point $k=0$ in the conduction bands can be lifted [see Fig. \ref{fig2}(d) and Fig. \ref{fig2}(f)] while the valence bands remain almost intact and look similar to the case $M = 0$ and $V \neq 0$ [see Fig. \ref{fig2}(c)]. Moreover, when $M > V$, the two minimums (maximums) of lower conduction (upper valence) band located at finite $k$ begin to merge together providing a single minimum (maximum) at $k=0$ [see Fig. \ref{fig2}(e)]. As a result, the band structure can be engineered via the combined effect of both $M$ and $V$ that can be exploited in the following.

%%%%%%%%%%%%%%%%%%%%%%%%%%%%%%%%%%%%%%%%%
\section{Pointlike impurities}\label{s3}
%%%%%%%%%%%%%%%%%%%%%%%%%%%%%%%%%%%%%%

In order to investigate the effect of disorder/impurity, we consider the potential of identical pointlike impurities in the form
\begin{equation}\label{imp}
V_{imp}(\mathbf{r})=u_0 \sigma_0\tau_0\sum_{i}\delta(\mathbf{r}-\mathbf{R}_{i}),
\end{equation}
where $u_0$ is the strength of scattering potential, $\delta(\mathbf{r})$ is the Dirac delta function, and $\mathbf{R}_{i}$ are the coordinates of randomly and equally distributed nonmagnetic impurities on the two surfaces of thin film. Also, $\sigma_0$ and $\tau_0$ are 2$\times$2 identity matrices in the spin and surface spaces, respectively. We assume that impurity correlations are Gaussian,
\begin{eqnarray}
\langle V_{imp}(\mathbf{r})\rangle
&=&0,\\
\langle V_{imp}(\mathbf{r}_1)V_{imp}(\mathbf{r}_2)\rangle&=&n_iu^2_0\delta(\mathbf{r}_1-\mathbf{r}_2),
\end{eqnarray}
where $n_i$ is the impurity density and $\langle\cdot\cdot\cdot\rangle$ is the average over space and impurity configurations.

We define the retarded and advanced disorder averaged Green functions as
\begin{equation}
G^{\pm}=\left[\mu-H-\Sigma^{\pm}\pm i \eta\right]^{-1},
\label{FullGreen}
\end{equation}
where $\Sigma^\pm$ are the self-energies, $\mu$ is the chemical potential, and $\eta\rightarrow0^+$. Using Dyson's series \cite{Dyson}, in the Born approximation, one can expand the impurity-averaged Green's function (\ref{FullGreen}) and gets
\begin{equation}
G^{\pm}=\frac{G_{0}^{\pm}}{1+\Sigma^{\pm}G_{0}^{\pm}},
\label{FullGreen1}
\end{equation}
where
\begin{equation}
G^{\pm}_0=\left[\mu-H\pm i \eta\right]^{-1},
\label{BareGreen}
\end{equation}
are the bare retarded and advanced Green's functions corresponding to the Hamiltonian \eqref{hamil}, that, in special cases, can be expressed as
\begin{equation}
\begin{aligned}
G_{\pm}^{0}=\frac{1}{S_{\pm}}[\chi_{\pm}\left(g+V\left(V+2 H_{R}\right)\right)-\Delta\left(g+V\left(V+2 H_{R}\right)\right) \tau_{x}
+\left(\tilde{V}-\left(g-V^{2}\right) H_{R}\right) \tau_{z}],
\label{GreenM}
\end{aligned}
\end{equation}
with
\begin{equation}
\begin{aligned}
S_{\pm}=\left(\Delta^{2}-\chi_{\pm}^{2}\right)\left(\left(\Delta^{2}-\chi_{\pm}^{2}\right)+2\left(V^{2}+v_{F k}^{2} k^{2}\right)\right)+\left(V^{2}-v_{F k}^{2} k^{2}\right)^{2},
\end{aligned}
\end{equation}
for $ V\neq 0$  and $M = 0$ and \begin{equation}
\begin{aligned}
G_0^{\pm}=\frac{1}{C_\pm}[\chi_{\pm} g-{\Delta}\left(g+M \chi_{\pm} \sigma_{z}\right) \tau_{x}+M {\Delta} H_{y} \tau_{y}-g H_{R} \tau_{z}-\lambda\tau_{+}+M \left( v_{F k}^{2} k^{2}-\chi_{\pm}^{2}\right) \sigma_{z} \tau_{-}],
\label{GreenV}
\end{aligned}
\end{equation}
with
\begin{equation}
\begin{aligned}
C_{\pm}=v_{F k}^{2} k^{2}\left(v_{F k}^{2} k^{2}+M^{2}-2\left(\chi_{\pm}^{2}-\Delta^{2}\right)\right)+\left(\chi_{\pm}^{2}-\Delta^{2}\right)^{2}-M \chi_{\pm}^{2},
\end{aligned}
\end{equation}
for $V = 0$  and $M \neq 0$ where
\begin{equation}
\begin{aligned}
&g=v_{F k}^{2} k^{2}+\Delta^{2}-\chi_{\pm}^{2}, \\
&\tilde{V}=V\left(V^{2}+\Delta^{2}-\chi_{\pm}^{2}-v_{F k}^{2} k^{2}\right),\\
&\chi_{\pm}=\left( r k^{2}-\mu \pm i \eta \right),\\
&H_{y}=v_{F k}\left(k_{x} \sigma_{x}+k_{y} \sigma_{y}\right),\\
&\lambda=M \left(M\left(\chi_{\pm}+ H_{R}\right) +\Delta^{2} \sigma_{z}\right).
\end{aligned}
\end{equation}
The self-energies $\Sigma^\pm$ can also be defined as \cite{selfEnergy}
\begin{equation}
\Sigma^{\pm}=\left\langle V_{\mathrm{imp}} G^{\pm} V_{\mathrm{imp}}\right\rangle.
\label{selfEn}
\end{equation}
The above relation, using Eq. (\ref{imp}), can be read as
\begin{equation}
\Sigma^{\pm}=n_{i} u_{0}^{2} \int\frac{d^{2} k}{(2\pi)^{2}} G^{\pm}.
\label{selfEn1}
\end{equation}
The self-energies $\Sigma^\pm$ can be written in the matrix structure,
\begin{equation}
\Sigma^{\pm}=\left(
  \begin{array}{cccc}
    \Sigma^{\pm}_{00} & \Sigma^{\pm}_{0x} & \Sigma^{\pm}_{0y} & \Sigma^{\pm}_{0z} \\
    \Sigma^{\pm}_{x0} & \Sigma^{\pm}_{xx} & \Sigma^{\pm}_{xy} & \Sigma^{\pm}_{xz}  \\
    \Sigma^{\pm}_{y0} & \Sigma^{\pm}_{yx} & \Sigma^{\pm}_{yy} & \Sigma^{\pm}_{yz}  \\
    \Sigma^{\pm}_{z0} & \Sigma^{\pm}_{zx} & \Sigma^{\pm}_{zy} & \Sigma^{\pm}_{zz}  \\
  \end{array}
\right),
\label{MetrixselfEn}
\end{equation}
where $\Sigma^{\pm}_{i,j}$ with $i,j=0,x,y,z$ are the elements of self-energies. Using \eqref{FullGreen1}, \eqref{BareGreen}, and \eqref{selfEn1}, it is easy to show that the nonzero elements are \cite{selfEnergy1,selfEnergy2}
\begin{equation}
\begin{aligned}
&\Sigma_{ii}^{\pm}=n_{i} u_{0}^{2} \int \frac{d^{2} k}{(2\pi)^{2}}  \frac{S_{i i}^{\pm}}{D_{+} D_{-}}, \quad i=0,x,y,z, \\
&\Sigma_{0 x}^{\pm}=n_{i} u_{0}^{2} \int \frac{d^{2} k}{(2\pi)^{2}}  \frac{S_{0 x}^{\pm}}{D_{+} D_{-}}, \\
&\Sigma_{y z}^{\pm}=n_{i} u_{0}^{2} \int \frac{d^{2} k}{(2\pi)^{2}}  \frac{S_{y z}^{\pm}}{D_{+} D_{-}},
\label{SCBA}
\end{aligned}
\end{equation}
where $D_{\pm}=Det(G^\pm)$ and $S^{\pm}_{i,j}$ with $i,j=0,x,y,z$ are given in the Supplemental Material \cite{Sup}. Here, the self-energies $\Sigma_{0 x}=\Sigma_{x 0}$ and $\Sigma_{y z}=\Sigma_{z y}$.
The self-energies \eqref{SCBA} can be calculated by the SCBA method. In the SCBA, with an initial guess of the self-energies $\Sigma^\pm$, one can determine a new value for them. This process can be done iteratively until the difference between successive values of the self-energy becomes smaller than a desired value. On the other hand, in the fBA, we replace $G^{\pm}$ by $G_0^{\pm}$ in Eq. \eqref{selfEn1}. The different components of self-energies in this method are given by
\begin{equation}
\begin{aligned}
&\Sigma_{00}^{\pm}=n_{i} u_{0}^{2} \int \frac{d k}{2\pi} k \frac{\chi_\pm \left(g+V^{2}\right)+\tilde{V}+M^{2}(\chi_\pm - V)-M \Delta^{2} }{\Delta^{4}+2 \Delta^{2}\left(V^{2}+v_{F k}^{2} k^{2}-\chi_\pm^{2}\right)+\left(v_{F k}^{2} k^{2}-(V-\chi_\pm)^{2}\right)\left(M^{2}+v_{F k}^{2} k^{2}-(V+\chi_\pm)^{2}\right)},\\
&\Sigma_{x x}^{\pm}=n_{i} u_{0}^{2} \int \frac{d k}{2\pi} k \frac{\chi_\pm \left(g+V^{2}\right)-\tilde{V}-\tilde{M} }{\Delta^{4}+2 \Delta^{2}\left(V^{2}+v_{F k}^{2} k^{2}-\chi_\pm^{2}\right)+\left(v_{F k}^{2} k^{2}-(V-\chi_\pm)^{2}\right)\left(M^{2}+v_{F k}^{2} k^{2}-(V+\chi_\pm)^{2}\right)}, \\
&\Sigma_{yy}^{\pm}=n_{i} u_{0}^{2} \int \frac{d k}{2\pi} k \frac{\chi_\pm \left(g+V^{2}\right)+\tilde{V}+M^{2}(\chi_\pm - V)+M \Delta^{2} }{\Delta^{4}+2 \Delta^{2}\left(V^{2}+v_{F k}^{2} k^{2}-\chi_\pm^{2}\right)+\left(v_{F k}^{2} k^{2}-(V-\chi_\pm)^{2}\right)\left(M^{2}+v_{F k}^{2} k^{2}-(V+\chi_\pm)^{2}\right)},\\
&\Sigma_{zz}^{\pm}=n_{i} u_{0}^{2} \int \frac{d k}{2\pi} k \frac{\chi_\pm \left(g+V^{2}\right)-\tilde{V}+\tilde{M} }{\Delta^{4}+2 \Delta^{2}\left(V^{2}+v_{F k}^{2} k^{2}-\chi_\pm^{2}\right)+\left(v_{F k}^{2} k^{2}-(V-\chi_\pm)^{2}\right)\left(M^{2}+v_{F k}^{2} k^{2}-(V+\chi_\pm)^{2}\right)}, \\
&\Sigma_{0 x}^{\pm}=\Sigma_{x 0}^{\pm}=-n_{i} u_{0}^{2} \int \frac{d k}{2\pi} k \frac{\Delta\left(g+V^{2}-M(\chi_\pm -V)\right)}{\Delta^{4}+2 \Delta^{2}\left(V^{2}+v_{F k}^{2} k^{2}-\chi_\pm^{2}\right)+\left(v_{F k}^{2} k^{2}-(V-\chi_\pm)^{2}\right)\left(M^{2}+v_{F k}^{2} k^{2}-(V+\chi_\pm)^{2}\right)}, \\
&\Sigma_{y z}^{\pm}=\Sigma_{z y}^{\pm}=-n_{i} u_{0}^{2} \int \frac{d k}{2\pi} k \frac{\Delta\left(g+V^{2}+M(\chi_\pm - V)\right)}{\Delta^{4}+2 \Delta^{2}\left(V^{2}+v_{F k}^{2} k^{2}-\chi_\pm^{2}\right)+\left(v_{F k}^{2} k^{2}-(V-\chi_\pm)^{2}\right)\left(M^{2}+v_{F k}^{2} k^{2}-(V+\chi_\pm)^{2}\right)}.
\label{fBA}
\end{aligned}
\end{equation}
Here, we have introduced
\begin{equation}
\tilde{M}=M\left(\chi_\pm^{2}+V^{2}-2 \chi_\pm V-v_{F k}^{2} k^{2}\right).
\end{equation}
For $r=0$, $|\mu| > \Delta$, and in the limit  $M\ll1$ and $V\ll1$, one can obtain short analytical approximated expressions to Eqs. (\ref{fBA}) as
\begin{equation}
\begin{aligned}
&\Sigma_{00}^{\pm}\approx  \frac{ \gamma_b}{\pi} \left(2 V-\frac{M \Delta^{2}}{\Delta^{2}-\mu^{2}}+(V-\mu) \log \left(\frac{\mu^{2}-\Delta^{2}}{\mu^{2}-\Delta^{2}+(k_c v_F)^2}\right) \pm  i\pi sgn(\mu)(\mu-V)\right), \\
&\Sigma_{x x}^{\pm}\approx \frac{\gamma_b}{\pi} \left(-2 V-\frac{M( \Delta^{2}-2V\mu)}{\Delta^{2}-\mu^{2}}-(M+V+\mu) \log \left(\frac{\mu^{2}-\Delta^{2}}{\mu^{2}-\Delta^{2}+(k_c v_F)^2}\right) \pm  i\pi sgn(\mu)(M+V+\mu)\right),\\
&\Sigma_{y y}^{\pm}\approx \frac{\gamma_b}{\pi} \left(2 V+\frac{M \Delta^{2}}{\Delta^{2}-\mu^{2}}+(V-\mu) \log \left(\frac{\mu^{2}-\Delta^{2}}{\mu^{2}-\Delta^{2}+(k_c v_F)^2}\right) \pm  i\pi sgn(\mu)(\mu-V)\right), \\
&\Sigma_{z z}^{\pm}\approx \frac{\gamma_b}{\pi} \left(-2 V+\frac{M( \Delta^{2}-2V\mu)}{\Delta^{2}-\mu^{2}}+(M-V-\mu) \log \left(\frac{\mu^{2}-\Delta^{2}}{\mu^{2}-\Delta^{2}+(k_c v_F)^2}\right) \pm  i\pi sgn(\mu)(-M+V+\mu)\right), \\
&\Sigma_{0 x}^{\pm}=\Sigma_{x 0}^{\pm}\approx  \frac{- \gamma_b}{\pi} \left(\frac{\Delta M(V-\mu)}{\left(\Delta^{2}-\mu^{2}\right)}-\Delta \log \left(\frac{\mu^{2}-\Delta^{2}}{\mu^{2}-\Delta^{2}+(k_c v_F)^2}\right) \pm i\pi sgn(\mu) \Delta\right), \\
&\Sigma_{y z}^{\pm}=\Sigma_{z y}^{\pm}\approx \frac{- \gamma_b}{\pi} \left(-\frac{\Delta M(V-\mu)}{\left(\Delta^{2}-\mu^{2}\right)}-\Delta \log \left(\frac{\mu^{2}-\Delta^{2}}{\mu^{2}-\Delta^{2}+(k_c v_F)^2}\right) \pm i\pi sgn(\mu) \Delta\right),
\end{aligned}
\label{ApproxfBA}
\end{equation}
where $k_c=\pi/a$ is the wave vector cutoff, $sgn(x)$ is the Sign function, and $\gamma_{b}=n_{i} u_{0}^{2} /\left(4 v_{F}^{2}\right)$ is the impurity parameter. In the gap region, $|\mu| < \Delta$, due to the absence of states, the self-energies take exponentially small values \cite{disorderedScatGap1,disorderedScatGap2}. Generally, the self-energies (\ref{SCBA}) can be written as
\begin{equation}
\Sigma^{\pm}=\Sigma^{\prime} \pm i \Gamma,
\end{equation}
where $\Sigma^{\prime}$ is real part and $\Gamma$ is imaginary part. This can also be seen explicitly from Eqs. (\ref{ApproxfBA}).
\begin{figure}[t]
\centering
\includegraphics[width=8.5cm]{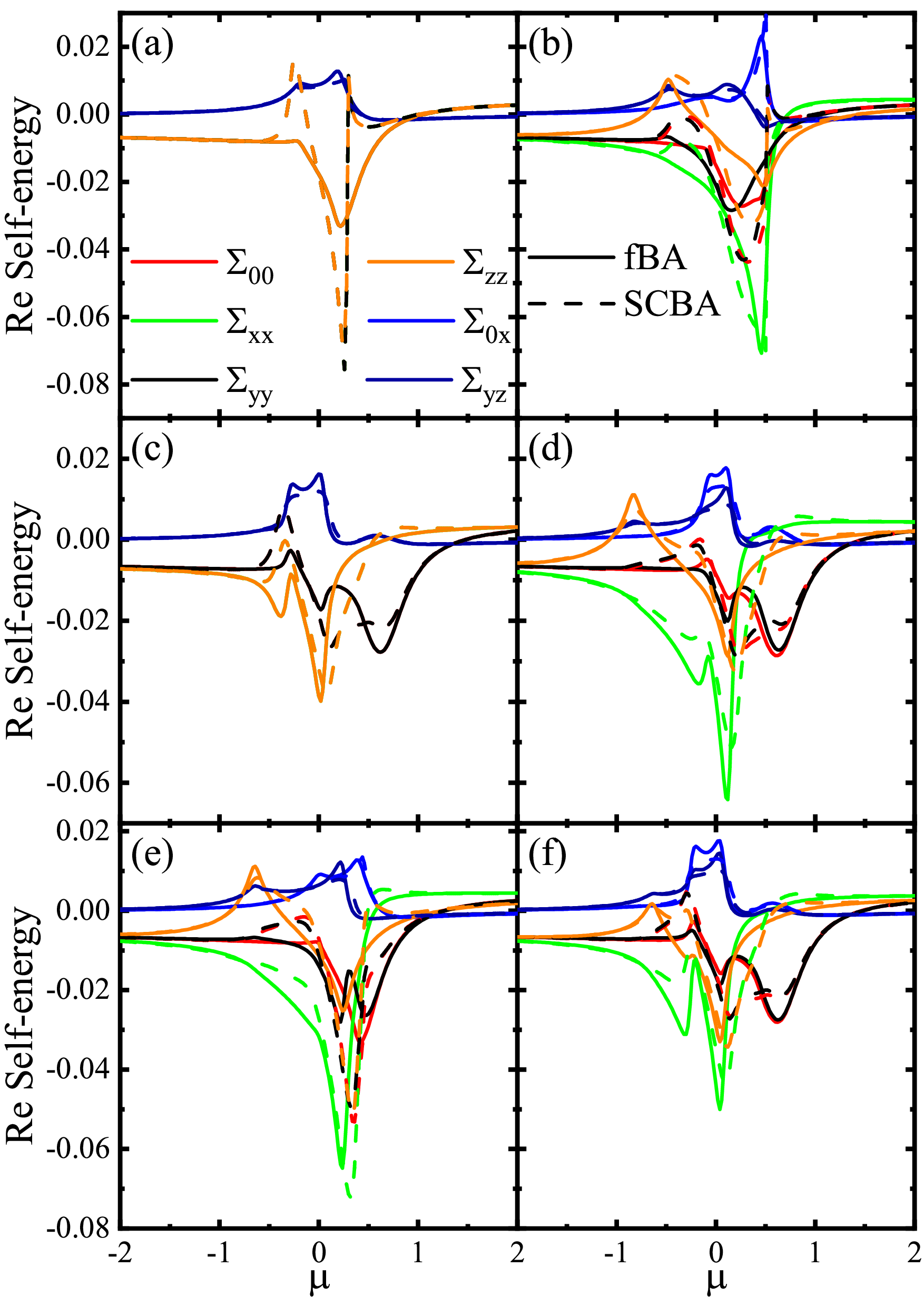}
\caption{(Color online) Real part of the self-energies as a function of the $\mu$ in the fBA and SCBA methods for (a) $M=0$ and $V=0$, (b) $M=0.4$ and $V=0$, (c) $M=0$ and $V=0.4$, (d) $M=0.4$ and $V=0.4$, (e) $M > V$ with $M=0.4$ and $V=0.2$, and (f) $M < V$ with $M=0.2$ and $V=0.4$. Here $\Delta = 0.2$, $\gamma_b=0.2$, $s=1$, and $r=1$.}
\label{reself}
\end{figure}
\begin{figure}[h]
\centering
\includegraphics[width=8.5cm]{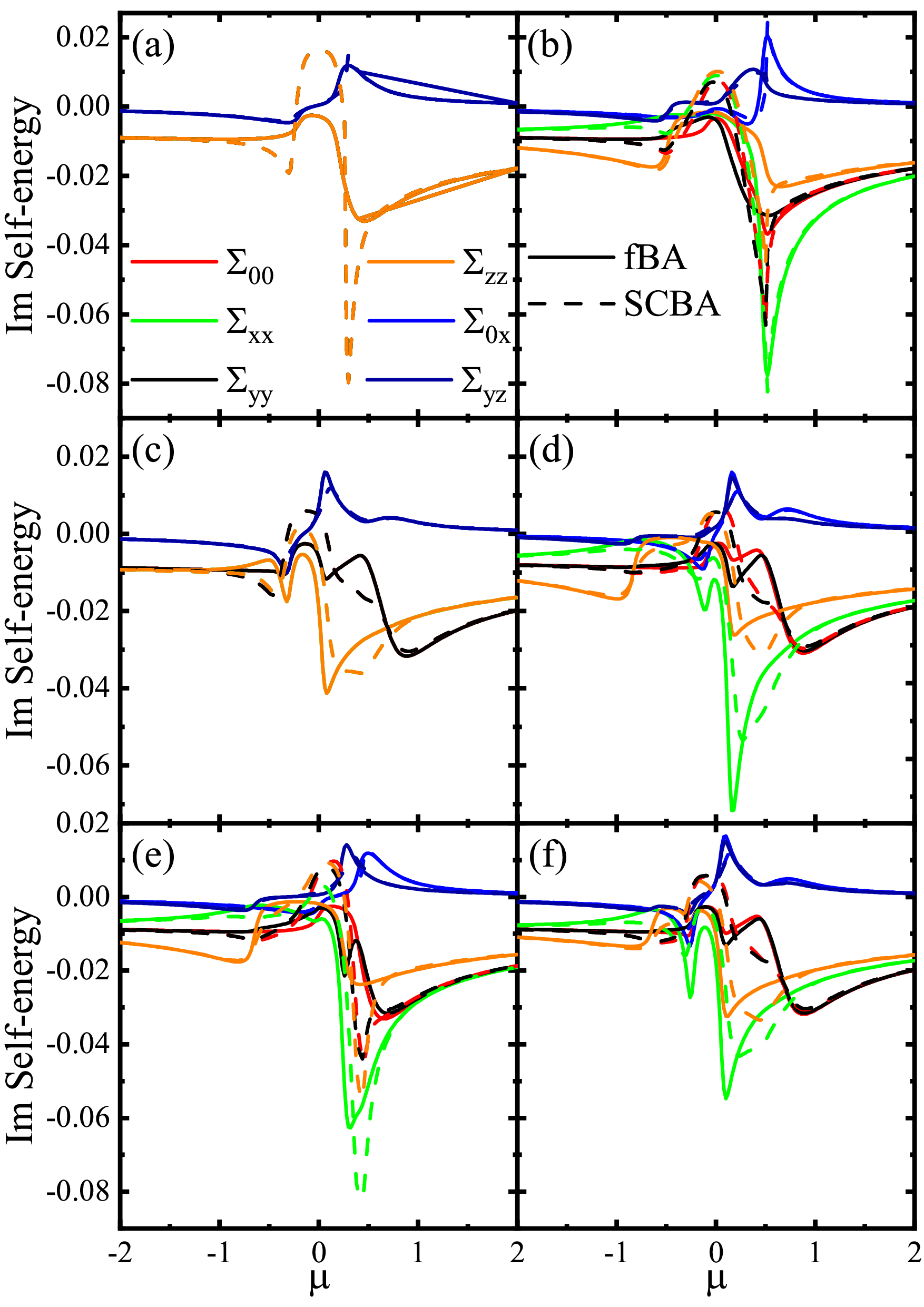}
\caption{(Color online) Imaginary part of the self-energies as a function of the $\mu$ in the fBA and SCBA methods for (a) $M=0$ and $V=0$, (b) $M=0.4$ and $V=0$, (c) $M=0$ and $V=0.4$, (d) $M=0.4$ and $V=0.4$, (e) $M > V$ with $M=0.4$ and $V=0.2$, and (f) $M < V$ with $M=0.2$ and $V=0.4$. Here $\Delta = 0.2$, $\gamma_b=0.2$, $s=1$, and $r=1$.}
\label{imself}
\end{figure}

The real and imaginary parts of self-energies versus $\mu$ with both the fBA and the SCBA methods are evaluated in Figs. \ref{reself} and \ref{imself} for different values of the $M$ and $V$. From both the figures, one can see that the obtained solutions of fBA usually correspond to those of SCBA method for relativity small $\gamma_b=0.2$. Both the real and imaginary parts of the self-energies change their values at low energies while at high energies often take small constant values. The diagonal components of self-energies are equal as well as the off-diagonal ones in the absence of $M$ and $V$, i.e., $\Sigma_{00}=\Sigma_{xx}=\Sigma_{yy}=\Sigma_{zz}$ and $\Sigma_{0x}=\Sigma_{yz}$ [see Figs. \ref{reself}(a) and \ref{imself}(a)]. This results from the presence of both time-reversal and inversion symmetries. Also, for $M=0$ and $V\neq0$ the components are $\Sigma_{xx}=\Sigma_{zz}$, $\Sigma_{00}=\Sigma_{yy}$, and $\Sigma_{0x}=\Sigma_{yz}$ [see Figs. \ref{reself}(c) and \ref{imself}(c)]. This is because of preserving time-reversal symmetry. On the other hand, for $M\neq0$ and $V=0$, the different components take different values [see Figs. \ref{reself}(b) and \ref{imself}(b)]. This indicates that, in our model, the $M$ breaks not only the time-reversal symmetry but also the inversion one. Furthermore, the real and imaginary parts of $\Sigma_{xx}$ usually gets large amplitudes compared with the others.

In the following, we study the disordered DOS \cite{disorderedDOS} using the Green's function technique. To do so, we start with
\begin{equation}
D=-\frac{1}{\pi} \int \frac{kdk}{2 \pi} \operatorname{Tr} \operatorname{Im}G^{+}.
\end{equation}
Plugging \eqref{FullGreen1} and \eqref{BareGreen} into the above equation, one gets
\begin{equation}
\begin{aligned}
D=\frac{-1}{\pi^{2}} \operatorname{Im} \int\frac{ k d k}{D_{+} D_{-}} [
M^{2}\left(\chi_{+}-V+i\left(\Gamma_{00}+\Gamma_{x x}\right)\right)
+2\left(V^{2}\!+\!v_{F k}^{2} \!k^{2}\!-\!\left(\chi_{+}\!+\!i\left(\Gamma_{y y}\!-\!\Gamma_{x x}\right)\right)^{2}\!-\!\left(\Gamma_{0 x}-i \Delta\right)^{2}\right)\left(\chi_{+}+i\left(\Gamma_{00}+\Gamma_{z z}\right)\right)].
\end{aligned}
\label{DOS1}
\end{equation}
For $r=s=0$, in the limit $\Delta \ll 1$ the above integral can be approximated as
\begin{equation}
\begin{aligned}
D \approx \frac{-2}{\pi^{2} k_{c}^{4} v_{F}^{6}} [5 \mu V^{2}\left(2 \Delta \Gamma_{0 x}-3 \mu\left(\Gamma_{00}+\Gamma_{x x}\right)\right)+k_{c}^{2} v_{F}^{2}\left(2 \Delta \mu \Gamma_{0 x}-3\left(\Gamma_{00}+\Gamma_{x x}\right)\left(V^{2}+\mu^{2}\right)\right)+2 k_{c}^{4} v_{F}^{4} \Gamma_{00} \log \left(k_{c}\right)],
\end{aligned}
\end{equation}
for $M=0$ and $V\neq0$ and
\begin{equation}
\begin{aligned}
D \approx \frac{-1}{\pi^{2} k_{c}^{4} v_{F}^{6}} [M^{2}\left(3 \mu^{2}\left(\tilde{\Gamma}_{1}+\tilde{\Gamma}_{2}\right)-4 \mu \Delta \Gamma_{0 x}\right)+k_{c}^{2} v_{F}^{2} \left(4 \mu \Delta \Gamma_{0 x}+(M^{2}-6 \mu^2)\left(\tilde{\Gamma}_{1}+\tilde{\Gamma}_{2}\right)\right)+4 k_{c}^{4} v_{F}^{4} \tilde{\Gamma}_{1} \log \left(k_{c}\right)],
\end{aligned}
\end{equation}
for $M\neq0$ and $V=0$, where $\tilde{\Gamma}_{1}=\Gamma_{00}+\Gamma_{y y}$ and $\tilde{\Gamma}_{2}=\Gamma_{x x}+\Gamma_{z z}$.

\begin{figure}[t]
\centering
\includegraphics[width=8.5cm]{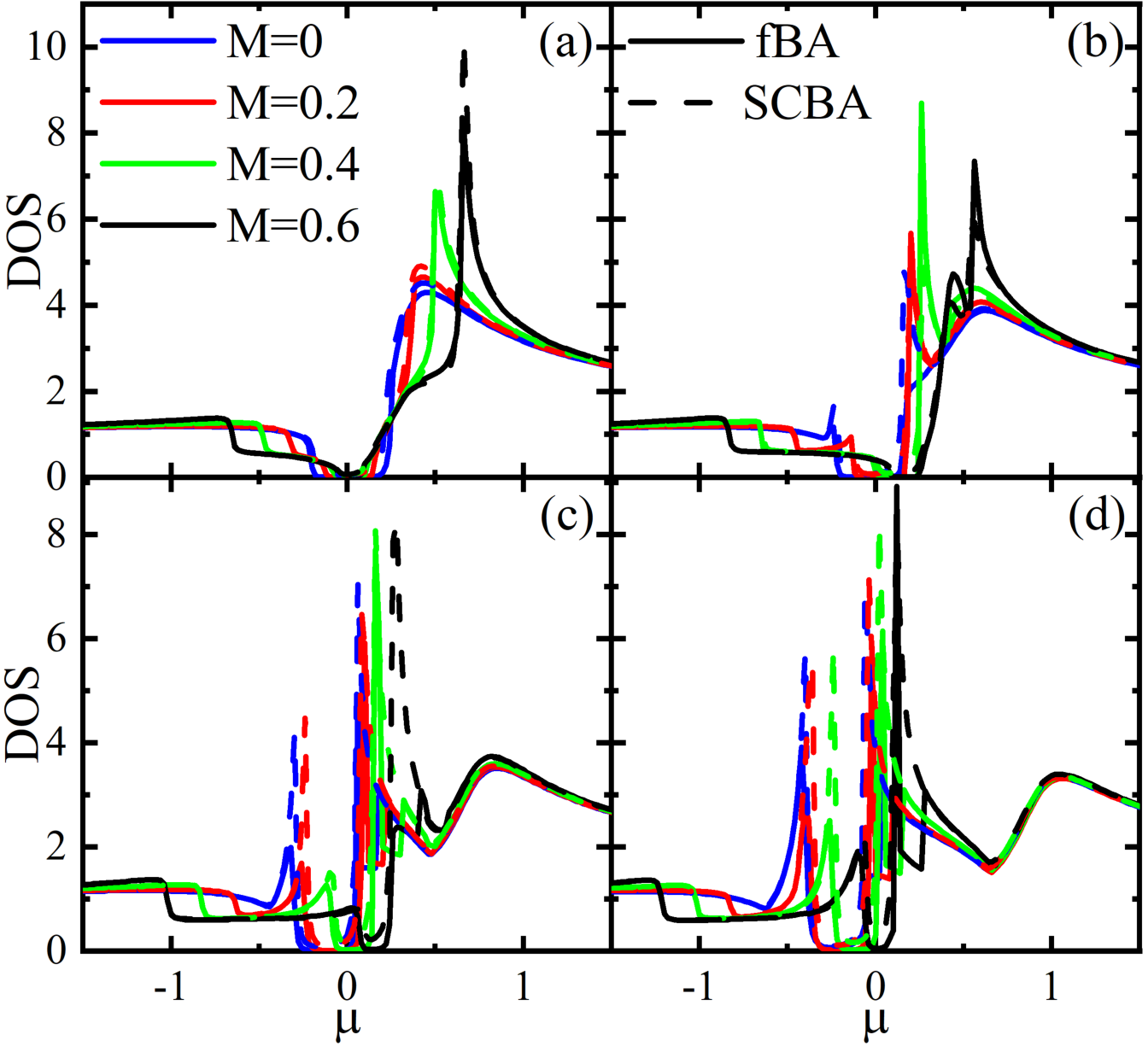}
\caption{(Color online) Dependence of DOS on the $\mu$ in the fBA and SCBA methods with $\gamma_{b}=0.2$ for different values of M and (a) $V=0$, (b) $V=0.2$, (c) $V=0.4$, and (d) $V=0.6$.}
\label{DOS}
\end{figure}
Numerical evaluation of \eqref{DOS1} as a function of the $\mu$ with different values of the $M$ and $V$ is plotted in Fig. \ref{DOS} for the fBA and SCBA. The DOSs of conduction and valence bands are not equal. The DOS of conduction band is larger than that of the valence band. As the $M$ increases, the DOS increases and, at the same time, the gap decreases. The gap will be vanished for large values of the $M$. For the small values of $V$ ($V=0$ and  $0.2)$, in the valence band, the DOS is almost constant and there is a maximum value of the DOS in the conduction band. With the increases of $V$ ($V=0.4$  and $0.6$), the coherent peaks at both the edges of gap become pronounced. Furthermore, for small $V$, the DOS calculated by the fBA and SCBA are almost the same [see Figs. \ref{DOS}(a) and \ref{DOS}(b)]. But, for large $V$, the SCBA results deviate from fBA ones around the gap [see Figs. \ref{DOS}(c) and \ref{DOS}(d)].

%%%%%%%%%%%%%%%%%%%%%%%%%%%%%%%%%%%%%%%%%%%%55
\section {Spin conductivity within Kubo formula}\label{s4}
%%%%%%%%%%%%%%%%%%%%%%%%%%%%%%%%%%%%%%%%%%%%%%%%%%%%5

In the linear response theory, a spin current can respond to an applied voltage through a spin conductivity that can be calculated by the Kubo formula \cite{Dyson}. The spin conductivity can be decomposed into three terms as \cite{spincond1,spincond2,spincond3}
\begin{equation}
\sigma_{\alpha \beta}^{\gamma}=\sigma_{\alpha \beta}^{I \gamma}+\sigma_{\alpha \beta}^{I I \gamma}+\sigma_{\alpha \beta}^{I I I \gamma},\end{equation}
where the first and second terms include the contribution of states at the Fermi level and the third one contains the contribution of states below the Fermi level. Here, the indices $\alpha$ and $\beta$ stand for the coordinates $x$ and $y$, and $\gamma$ indicates the direction of spin.

The spin conductivity due to Fermi surface states, at zero temperature, can be written as \cite{spincond1}
\begin{equation}
\sigma_{\alpha \beta}^{I \gamma}=\frac{e}{4 \pi} \int \frac{d^{2} k}{(2 \pi)^{2}} \operatorname{Tr}\left[j_{\alpha}^{\gamma} G^{+} \tilde{v}_{\beta} G^{-}\right],
\label{sc1}
\end{equation}
\begin{equation}
\sigma_{\alpha \beta}^{I I \gamma}=-\frac{e}{8 \pi} \int \frac{d^{2} k}{(2 \pi)^{2}} \operatorname{Tr}\left[j_{\alpha}^{\gamma} G^{+} \tilde{v}_{\beta} G^{+}+j_{\alpha}^{\gamma} G^{-} \tilde{v}_{\beta} G^{-}\right],
\label{sc2}
\end{equation}
where $j_{\alpha}^{\gamma}=\left\{\sigma_{\gamma}, v_{\alpha}\right\}/4$ is the current operator with $v_{\alpha}=\partial H / \partial k_{\alpha}$ being the velocity operator, $\tilde{v}_{\alpha}$ is the velocity-vertex function. Note that the terms including only retarded or advanced Green's functions, i.e., $\left<j_{\alpha}^{\gamma} G^{+} v_{\beta} G^{+}\right>$ or $\left<j_{\alpha}^{\gamma} G^{-} v_{\beta} G^{-}\right>$, can be neglected in the weak scattering limit \cite{spincond2,zerothterm0,zerothterm1}. The spin conductivity of filled states below the Fermi level is given by \cite{spincond1,spincond2,spincond3}
\begin{equation}
\begin{aligned}
\sigma_{\alpha \beta}^{I I I \gamma}= \frac{e}{8 \pi} \int \frac{d^{2} k}{(2 \pi)^{2}} \int_{-\infty}^{\mu} f(E) d E \operatorname{Tr}\left[j_{\alpha}^{\gamma} G^{+}  \tilde{v}_{\beta} \frac{d G^{+}}{d E}-j_{\alpha}^{\gamma} \frac{d G^{+}}{d E}  \tilde{v}_{\beta} G^{+}+\mathrm{c.c.}\right],
\end{aligned}
\label{sc3}
\end{equation}
where $f(E)$ is the Fermi-Dirac function, and c.c denotes complex conjugate. We first neglect the effect of vertex corrections, in the next section, their effect will be investigated.

In the SCBA, having obtained the self-energies self-consistently and the impurity-averaged Green's function, we put $G^{\pm}$ into Eqs. \eqref{sc1}-\eqref{sc3} and calculate the spin conductivities. But in the fBA, we replace $G^{\pm}$ by $G_0^{\pm}$ in Eq. \eqref{selfEn1} and substitute the self-energies \eqref{selfEn1} into Eqs. \eqref{sc1}-\eqref{sc3} and calculate the spin conductivities. Note that for Hamiltonian \eqref{hamil}, although the spin conductivity tensor $\sigma_{\alpha \beta}^{I I I \gamma}$ can take finite values in the special case $M=V=0$ \cite{spincond3}, but it vanishes in the cases ($M=0, V\neq0$), ($M\neq0, V=0$), or ($M\neq0, V\neq0$), due to breaking the inversion symmetry. Then the term $\sigma_{\alpha \beta}^{I \gamma}$ is the only considerable one that survives in the system. To be more specific, in what follows, we focus on the $z$-component of transverse spin conductivity. Using Eq. \eqref{sc1} and employing Eqs. \eqref{FullGreen1}, \eqref{GreenM}, and \eqref{GreenV}, the spin conductivity $\sigma_{x y}^{I z}$ in the limit $\Delta \ll 1$, $r \ll 1$, and $s=0$ can be approximated as
\begin{equation}
\begin{aligned}
\sigma_{x y}^{I z} \approx \frac{4 \sigma_{0}^{z} \pi r}{k_{c}^{4} v_{F}^{6}}[V^{2} \mu\left(9 \mu\left(\Gamma_{00}+\Gamma_{x x}\right)-4 \Delta \Gamma_{0 x}\right)+2 k_{c}^{2} v_{F}^{2}\left(\Gamma_{00}+\Gamma_{x x}\right)\left(V^{2}+\mu^{2}\right)-2 k_{c}^{4} v_{F}^{4} \log \left(k_{c}\right)],
\end{aligned}
\end{equation}
for $M=0$ and $V\neq0$ and
\begin{equation}
\begin{aligned}
\sigma_{x y}^{I z} \approx-\frac{2 \sigma_{0}^{z} \pi r}{k_{c}^{4} v_{F}^{6}}[M^{2} \tilde{\Gamma}_{2}\left(2 k_{c}^{2} v_{F}^{2}+3 \mu^{2}\right)+2 k_{c}^{2} v_{F}^{2}\left(\tilde{\Gamma}_{1}+\tilde{\Gamma}_{2}\right)\left(k_{c}^{2} v_{F}^{2} \log \left(k_{c}\right)-\mu^{2}\right)],
\end{aligned}
\end{equation}
for $M\neq0$ and $V=0$, where $\sigma_{0}^{z}=e / 16 \pi^{3}$.
\begin{figure}[t]
\centering
\includegraphics[width=8.5cm]{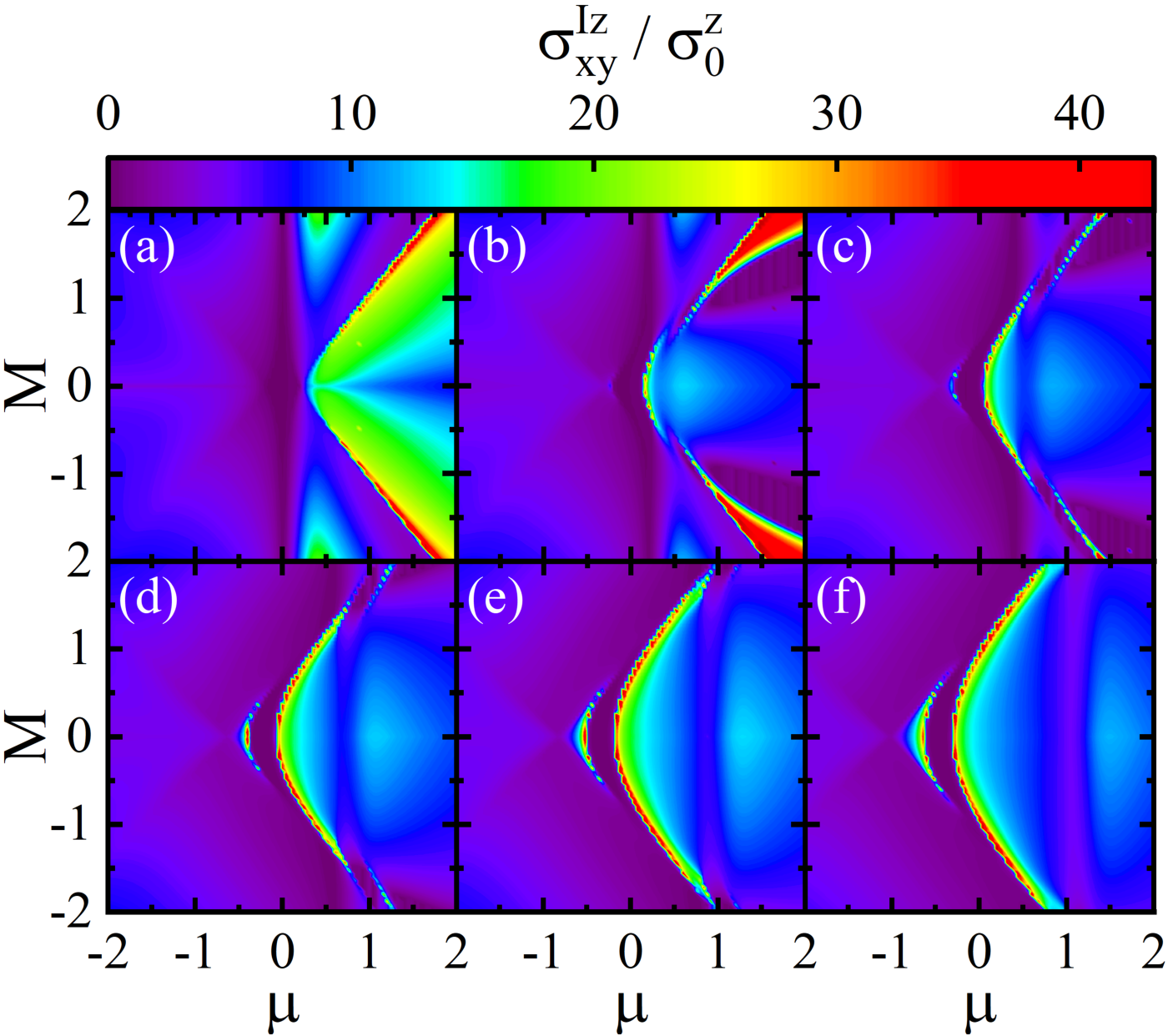}
\caption{(Color online) Density plot of the spin conductivity $\sigma_{x y}^{I z}$ in the fBA method as functions of the $M$ and $\mu$ with the disorder parameter $\gamma_{b}=0.2$ for (a) $V=0$, (b) $V=0.2$, (c) $V=0.4$, (d) $V=0.6$, (e) $V=0.8$, and (f) $V=1$.}
\label{DensityM}
\end{figure}

In Fig. \ref{DensityM}, the density plot of the spin conductivity $\sigma_{x y}^{I z}$ as functions of the $M$ and $\mu$ is depicted in the fBA for various values of the $V$. All plots show that the spin conductivity via the conduction band states can have larger values than those of the valence band arising from the large DOS in the conduction band and more dispersive feature of the conduction band. For $V=0$, as shown in Fig. \ref{DensityM}(a), there is a global gap around $\mu=0$ with the largest bandgap width at $M=0$. Also, a large spin conductivity occurs for large $|M|$ in the conduction band away the charge neutrality point, $\mu=0$. Interestingly, as shown in Fig. \ref{DensityM}(b) for $V=0.2$ a region with large values of the spin conductivity splits into two parts so that one part shifts towards $\mu=0$ and the other one shifts oppositely as $V$ increases, see Figs. \ref{DensityM}(c)-\ref{DensityM}(f). At the same time, a considerable spin conductivity emerges in the top of valence band for small $M$. As a result, this changes the global gap into the partial gap and a large spin conductivity can take place at low dopings with small amplitudes of the $M$.

\begin{figure}[h]
\centering
\includegraphics[width=8.5cm]{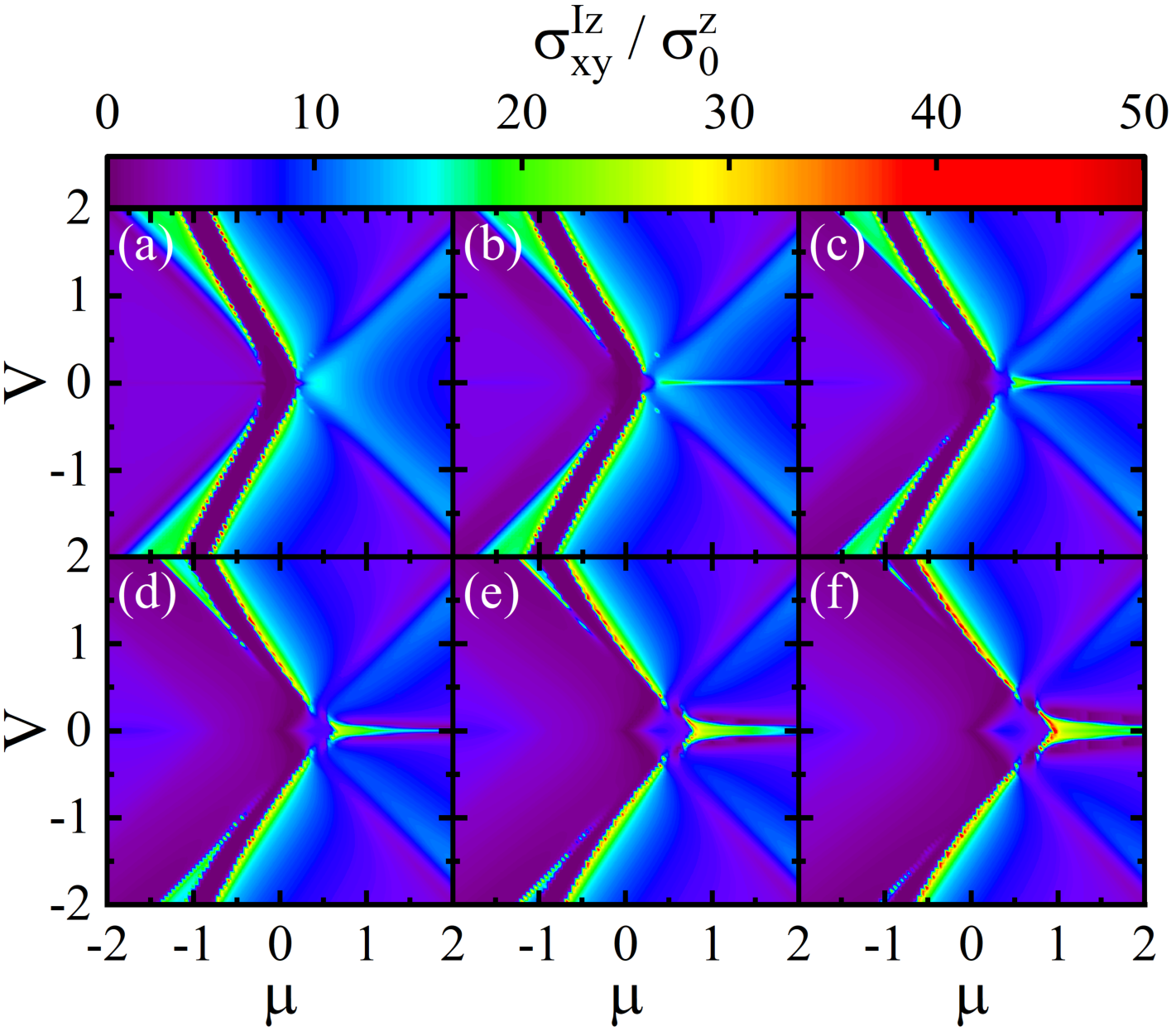}
\caption{(Color online) Density plot of the spin conductivity $\sigma_{x y}^{I z}$ in the fBA method as functions of the $V$ and $\mu$ with the disorder parameter $\gamma_{b}=0.2$ for (a) $M=0$, (b) $M=0.2$, (c) $M=0.4$, (d) $M=0.6$, (e) $M=0.8$, and (f) $M=1$.}
\label{DensityV}
\end{figure}

The spin conductivity density plot as functions of $V$ and $\mu$ is depicted for various values of the $M$ in Fig. \ref{DensityV}. For $M=0$, as can be seen from Fig. \ref{DensityV}(a), there is a partial gap between conduction and valence bands and the spin conductivity $\sigma_{x y}^{I z}$ takes moderate values near the edges of the gap with smallest values near $V=0$. As $M$ increases, the gap decreases, in particular, for small $V$ and the spin conductivity $\sigma_{x y}^{I z}$ near the lower edge of the gap begin to vanish [see Figs. \ref{DensityV}(b)-\ref{DensityV}(f)]. Moreover, the spin conductivity $\sigma_{x y}^{I z}$ increases not only near the upper edge of gap but also, interestingly, at high dopings for small values of the $V$.

\begin{figure}[t]
\centering
\includegraphics[width=8.5cm]{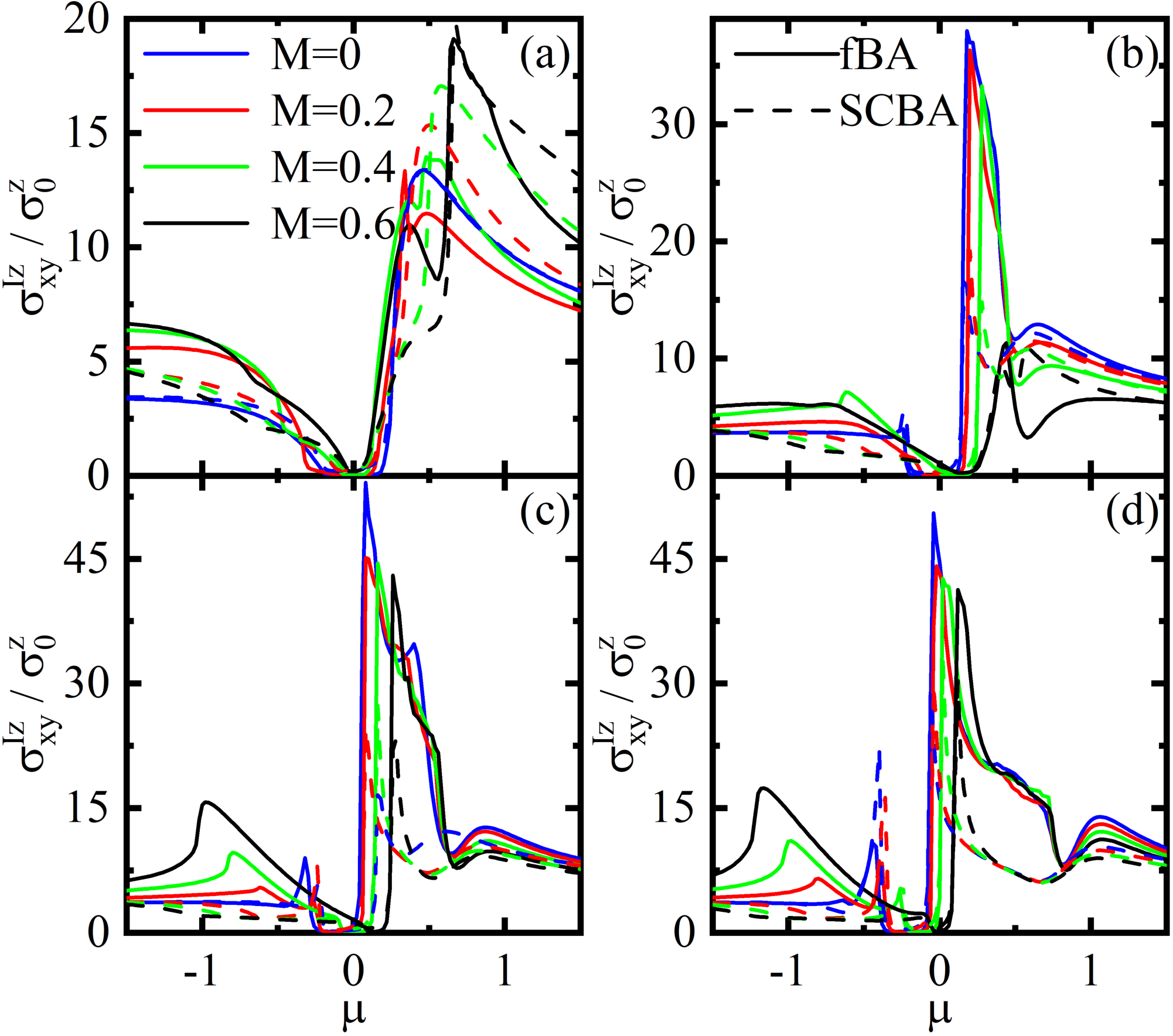}
\caption{(Color online) The spin conductivity $\sigma_{x y}^{I z}$ in the fBA and SCBA methods as a function of the $\mu$ for different values of M with (a) $V=0$, (b) $V=0.2$, (c) $V=0.4$, and (d) $V=0.6$. Here, the disorder parameter $\gamma_{b}=0.2$.}
\label{SpinConfBASCBA}
\end{figure}

In Fig. \ref{SpinConfBASCBA}, the spin conductivity $\sigma_{x y}^{Iz}$ is depicted in terms of the $\mu$ for various values of the $M$ and $V$ in both the fBA and SCBA methods. For $V=0$, see Fig. \ref{SpinConfBASCBA}(a), with $M=0$, for small impurity parameter $\gamma_{b}=0.2$ the fBA and SCBA have the same diagrams. Moreover, with the increase of $M$, the solutions of SCBA slightly deviate from those of the fBA. This is because of lifting the degeneracy at low energies, as already discussed. As $V$ increases, asymmetric coherent peaks at the edges of band gap appear, see Figs. \ref{SpinConfBASCBA}(b)-\ref{SpinConfBASCBA}(d). One also realizes that the deviation between both solutions becomes more pronounced at low energies. This can be attributed to lifting the degeneracy of more states when the $V$ turns on providing more available states required for the electron scattering off impurities at low energies. Subsequently, multi-scattering processes can come into play and higher order terms in the BA should be taken into account. This results in the difference between the fBA and SCBA results. In particular, the difference of both the solutions in the conduction band is considerable than that in the valence band. Because, the Fermi surfaces of conduction band is smaller than those of valence band increasing more scattering processes. Furthermore, at high energies, the difference between the solutions of the fBA and SCBA decreases, because the impurities affect on low energy states.

\begin{figure}[t]
\centering
\includegraphics[width=8.5cm]{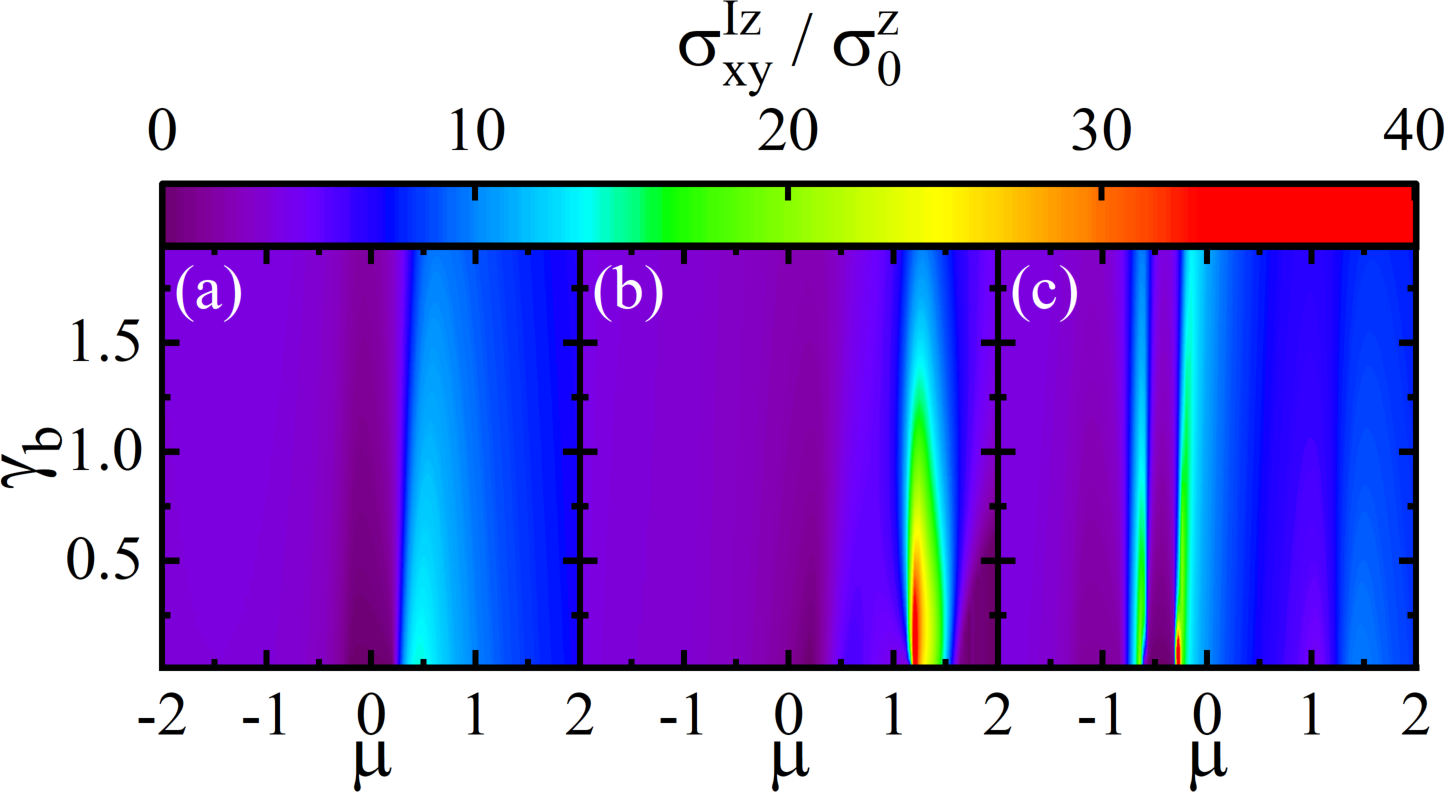}
\caption{(Color online) Density plot of the spin conductivity $\sigma_{xy}^{I z}$ in the SCBA method as functions of the disorder parameter $\gamma_{b}$ and $\mu$ for (a) $M=0$ and $V=0$, (b) $M > V$ with $M=1.5$ and $V=0.2$, and (c) $M < V$ with $M=0.2$ and $V=1$. Here $\Delta = 0.2$, s=1, and r=1.}
\label{ImpStre}
\end{figure}
The dependence of the spin conductivity $\sigma_{xy}^{I z}$ on the disorder parameter $\gamma_{b}$ and $\mu$ in the SCBA method is shown in Fig. \ref{ImpStre} for different values of the $M$ and $V$. In the absence of magnetic exchange field and potential difference, i.e., $M=0$ and $V=0$, the small values of spin conductivity  spoil as $\gamma_{b}$ increases [see Fig. \ref{ImpStre}(a)]. But, interestingly, for either $M > V$ with large $\mu$ [see Fig. \ref{ImpStre}(b)] or $M < V$ with small $\mu$ [see Fig. \ref{ImpStre}(c)], the obtained high spin conductivity can sustain even in the large values of $\gamma_{b}$ providing a stable spin Hall conductivity. Note that we have calculated the $\sigma_{x y}^{IIz}$ versus the disorder parameter $\gamma_{b}$ and $\mu$ (not shown). Its values is negligible compared to the $\sigma_{x y}^{Iz}$. So, the patterns of Fig. \ref{ImpStre} do not change by including the $\sigma_{x y}^{IIz}$.
\begin{figure}[h]
\centering
\includegraphics[width=8.5cm]{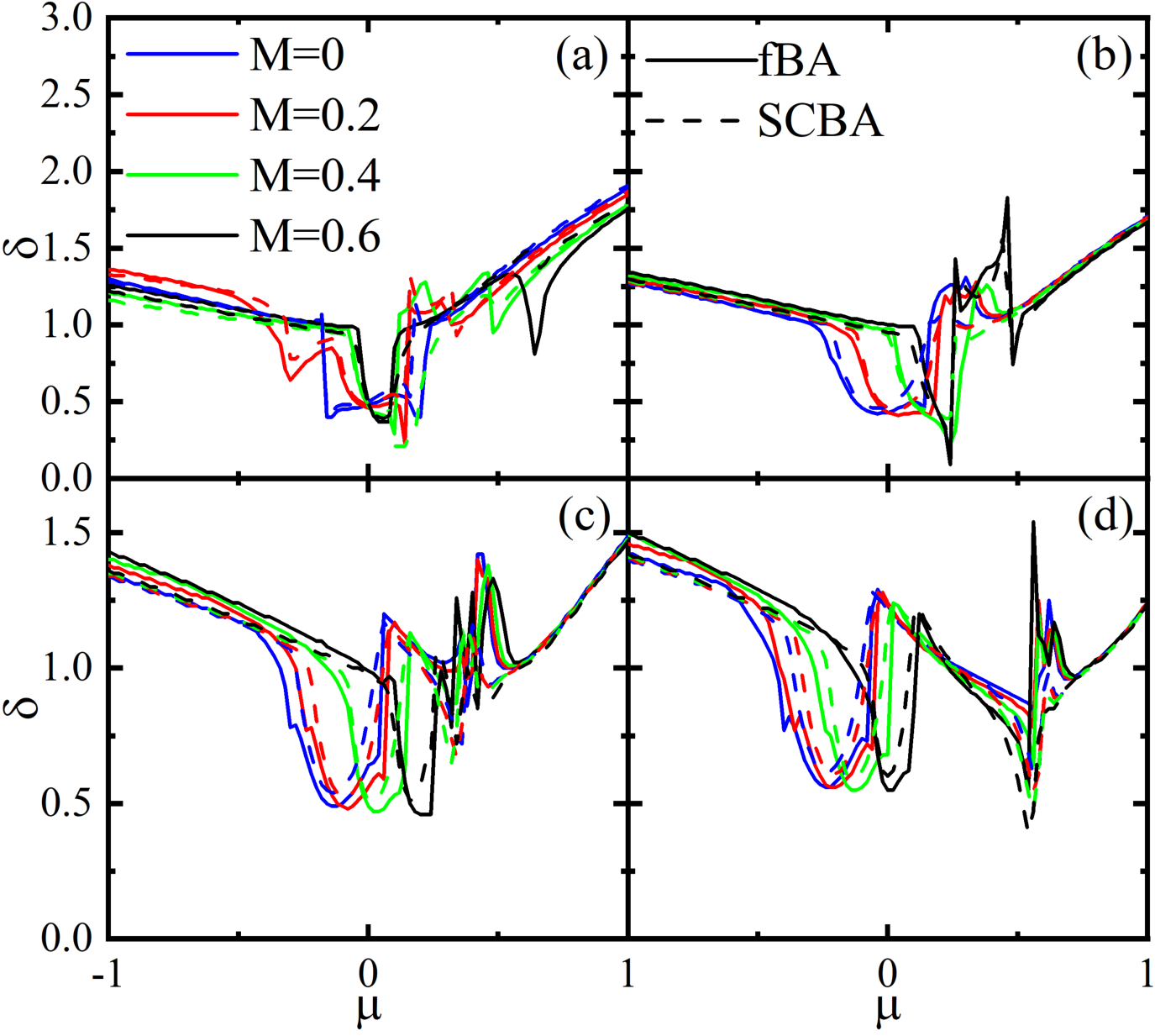}
\caption{\small (Color online) The vertex correction $\delta$ in the fBA and SCBA methods as a function of the $\mu$ for different values of $M$ with (a) $V=0$, (b) $V=0.2$, (c) $V=0.4$, and (d) $V=0.6$. Here, the disorder parameter is $\gamma_{b}=0.2$.}
\label{VCor}
\end{figure}
%%%%%%%%%%%%%%%%%%%%%%%%%%%%%%%%%%%%%%%%%%%%%555
\section {Vertex corrections}\label{s5}
%%%%%%%%%%%%%%%%%%%%%%%%%%%%%%%%%%%%%%%%5
The velocity-vertex function $\tilde{v}_{\alpha}$ satisfies the self-consistent equation \cite{zerothterm0,zerothterm1}
\begin{equation}
\tilde{v}_{\alpha} = v_{\alpha}+ n_{i}u_{0}^{2}\int \frac{d^{2}k}{(2\pi
)^{2}}G^{R}\tilde{v}_{\alpha}G^{A}.
\label{VeloVertex}
\end{equation}
To ensure about the true form of vertex function matrix structure, we calculate the first-order correction of velocity-vertex function $\tilde{v}_{\alpha}$. Using iteration, the first-order correction to velocity-vertex function reads
\begin{equation}
\tilde{v}_{\alpha}^{(1)} =  n_{i}u_{0}^{2}\int \frac{d^{2}k}{(2\pi
)^{2}}G^{R}v_{\alpha}G^{A},
\label{VereqnkR3}
\end{equation}
in the case $M=0$ and $V\ll1$ for $r=s=0$, it can be obtained as
\begin{equation}
\begin{aligned}
\tilde{v}_{x}^{(1)}=\frac{\gamma_{b} v_{F}}{\pi}[&\left(\frac{8 \Gamma_{x x} \Gamma_{0 x} \Delta \mu}{3\left(\Delta^{2}-\mu^{2}\right)^{2}}-1\right) \sigma_{y} \tau_{z}+\frac{2\left(\Gamma_{0 x} \mu+\Gamma_{00} \Delta\right)}{3\left(\Delta^{2}-\mu^{2}\right)} \sigma_{y} \tau_{y}+\frac{4 V \Gamma_{00} \Gamma_{0 x}\left(\Delta \sigma_{y} \tau_{0}-\mu \sigma_{y} \tau_{x}\right)}{3\left(\Delta^{2}-\mu^{2}\right)^{2}}],
\label{VeloVertex1}
\end{aligned}
\end{equation}

\begin{equation}
\begin{aligned}
\tilde{v}_{y}^{(1)}=-\frac{\gamma_{b} v_{F}}{\pi}[&\left(\frac{8 \Gamma_{x x} \Gamma_{0 x} \Delta \mu}{3\left(\Delta^{2}-\mu^{2}\right)^{2}}-1\right) \sigma_{x} \tau_{z}+\frac{2\left(\Gamma_{0 x} \mu+\Gamma_{00} \Delta\right)}{3\left(\Delta^{2}-\mu^{2}\right)} \sigma_{x} \tau_{y}+\frac{4 V \Gamma_{00} \Gamma_{0 x}\left(\Delta \sigma_{x} \tau_{0}-\mu \sigma_{x} \tau_{x}\right)}{3\left(\Delta^{2}-\mu^{2}\right)^{2}}].
\label{VeloVertex2}
\end{aligned}
\end{equation}
In the equations above, for weak scattering limit the first terms will be dominated, so one can expand the velocity vertex as
\begin{equation}
\begin{aligned}
&\tilde{v}_{x}=v_{x}+\delta \sigma_{y} \tau_{z}, \\
&\tilde{v}_{y}=v_{y}-\delta \sigma_{x} \tau_{z}.
\label{VeloVertex1}
\end{aligned}
\end{equation}
Plugging Eqs. (\ref{VeloVertex1}) into Eq. (\ref{VeloVertex}), one can obtain the correction $\delta$ to the velocities numerically. Note that for the case $V=0$ and $M\ll1$, we examined that the matrix structure of Eqs. (\ref{VeloVertex1}) remains the same.

In Fig. \ref{VCor}, the velocity correction $\delta$ is evaluated as a function of the $\mu$ in both the fBA and SCBA methods with different values of the $M$ and $V$. At high energies, the velocity correction $\delta$ has larger values than those for low energies almost independent of the $M$. Moreover, at low energies, there is a dip such that the wide of dip becomes narrowed as $M$ increases. For $V=0$, the dips are centered at $\mu=0$ as shown in Fig. \ref{VCor}(a). Also, as $V$ increases the center of dips deviates from zero depending on the $M$ and, at the same time, the $\delta$ exhibits a strong sudden change in $\mu>0$ as shown in Figs. \ref{VCor}(b)-\ref{VCor}(d). Note that although, unlike the spin Hall conductivity, the vertex correction to velocity is not an experimental observable but as can be seen from Eqs. \ref{sc1} and \ref{VeloVertex}, it influences the spin Hall conductivity being proportional to the difference between spin Hall conductivity with and without vertex correction.

%%%%%%%%%%%%%%%%%%%%%%%%%%%%%%%%%%%%%%%%%%%%%%%%%%%%%%%%%%%%%%%%%%%%%%%%%%%
\section {Summary} \label{s6}
%%%%%%%%%%%%%%%%%%%%%%%%%%%%%%%%%%%%%%%%%%%%%%%%%%%%%%%%%%%%%%%%%%%%%%%%%%%
We considered a disordered 2D TI thin film, having two surface states, subjected to a magnetic exchange field and a potential difference. The magnetic exchange field is applied perpendicular to a one of the surface states and the potential difference is between the two surfaces. Using the fBA and SCBA, we calculated both self-energies and DOS analytically and numerically. It is found that, in the model, a large spin Hall conductivity can be reached for either large magnetic exchange fields and small potential differences at high dopings or small magnetic exchange fields and large potential differences at low dopings. The promoted spin Hall effect is also examined with respect to the impurity parameter exhibiting a good stability against scattering of charge impurities. Also, the effect of vertex corrections is investigated changing the velocity at low energies.

Note that the magnetic gap formation in both magnetically doped TIs and a proximity-coupled magnetic insulator to TIs is a controversial task. In the former case, the gap depends on the impurity type and its location from the surface \cite{DopgapSize3}. While in the latter case, an electric gate can affect the interface magnetism \cite{Gategap}. The magnetic insulator film EuS, being coupled to the Bi$_2$Se$_3$ TI can provide a magnetic field with gap size 9 meV \cite{EffecHamil3, ProxgapSize2}. Also, the heterostructure MnBi$_2$Se$_4$/Bi$_2$Se$_3$ reveals ferromagnetism with gap of $\sim$ 100 meV \cite{ProxgapSize3}. Using  the realistic values $v_f = 3.5$ eV ${\AA}$, $a=4.19$ ${\AA}$, $\Delta\sim 10-100$ meV, $r=-19.5$ eV ${\AA}^2$, and $s=21$ ${\AA}^2$ for Bi$_2$Se$_3$ \cite{Realistic}, we find that the spin conductivity is about 5.6 $e^2/\hslash$ which is comparable with experiments in the time-reversal symmetry broken TI Mn$_x$Bi$_{2-x}$Te$_{3-y}$Se$_y$ \cite{Exp}.

\section*{Data availability}
All data generated or analyzed during this study are included in this published article and its supplementary
information files.

\section*{Author contributions}
M. V. Hosseini conceived the idea of the research and directed the project. All authors developed the research conceptions, analysed, discussed the obtained results, and wrote the paper. S. Pooyan performed the calculations.

\section*{Additional information}
\textbf{Competing financial interests:} The authors declare no competing financial and non-financial interests.

%%%%%%%%%% Merge with supplemental materials %%%%%%%%%%

%\clearpage
%\fontsize{10}{11}\selectfont
\clearpage
%%%%%%%%%% Merge with supplemental materials %%%%%%%%%%
\pagebreak
%\widetext
\begin{center}
\textbf{ Supplementary Material for Enhanced and stable spin Hall conductivity in a disordered time-reversal and inversion symmetry broken topological insulator thin film}
\end{center}

\begin{center}
Siamak Pooyan and Mir Vahid Hosseini\\
Department of Physics, Faculty of Science, University of Zanjan, Zanjan 45371-38791, Iran
\end{center}

\setcounter{equation}{0}
\setcounter{figure}{0}
\setcounter{table}{0}
\setcounter{page}{1}
\makeatletter
\renewcommand{\theequation}{S\arabic{equation}}
\renewcommand{\thefigure}{S\arabic{figure}}

\section*{Some notations} \label{AppxA}
We have introduced $S^{\pm}_{i,j}$ in Eq. (19) of the main text as

\begin{equation}
\begin{aligned}
S_{00}^{\pm}=&V\left(V^{2}-\left(v_{F k}^{2} k^{2}+\left(\chi_{\pm}+\Sigma_{00}\right)^{2}-\left(\Delta+\Sigma_{0 x}\right)^{2}\right)\right)-M\left(\left(V-\chi_{\pm}-\Sigma_{y y}\right) + \left(\Delta+\Sigma_{0 x}\right)^{2}\right) \\
&+\left(\chi_{\pm}+\Sigma_{z z}\right)\left(V^{2}-\left(-v_{F k}^{2} k^{2}+\left(\chi_{\pm}+\Sigma_{x x}\right)^{2}-\left(\Delta+\Sigma_{y z}\right)^{2}\right)\right),\\
S_{x x}^{\pm}=&-V\left(V^{2}-\left(v_{F k}^{2} k^{2}+\left(\chi_{\pm}+\Sigma_{00}\right)^{2}-\left(\Delta+\Sigma_{0 x}\right)^{2}\right)\right)+M\left(v_{F k}^{2} k^{2}+\left(V-\chi_{\pm}-\Sigma_{y y}\right)\right) \\
&+\left(\chi_{\pm}+\Sigma_{z z}\right)\left(V^{2}-\left(-v_{F k}^{2} k^{2}+\left(\chi_{\pm}+\Sigma_{y y}\right)^{2}-\left(\Delta+\Sigma_{y z}\right)^{2}\right)\right),\\
S_{yy}^{\pm}=&V\left(V^{2}-\left(v_{F k}^{2} k^{2}+\left(\chi_{\pm}+\Sigma_{00}\right)^{2}-\left(\Delta+\Sigma_{0 x}\right)^{2}\right)\right)-M\left(\left(V-\chi_{\pm}-\Sigma_{y y}\right) - \left(\Delta+\Sigma_{0 x}\right)^{2}\right) \\
&+\left(\chi_{\pm}+\Sigma_{z z}\right)\left(V^{2}-\left(-v_{F k}^{2} k^{2}+\left(\chi_{\pm}+\Sigma_{x x}\right)^{2}-\left(\Delta+\Sigma_{y z}\right)^{2}\right)\right),\\
S_{z z}^{\pm}=&-V\left(V^{2}-\left(v_{F k}^{2} k^{2}+\left(\chi_{\pm}+\Sigma_{00}\right)^{2}-\left(\Delta+\Sigma_{0 x}\right)^{2}\right)\right)-M\left(v_{F k}^{2} k^{2}-\left(V-\chi_{\pm}-\Sigma_{y y}\right)\right) \\
&+\left(\chi_{\pm}+\Sigma_{z z}\right)\left(V^{2}-\left(-v_{F k}^{2} k^{2}+\left(\chi_{\pm}+\Sigma_{y y}\right)^{2}-\left(\Delta+\Sigma_{y z}\right)^{2}\right)\right),\\
S_{0 x}^{\pm}=&-\left(\Delta+\Sigma_{0x}\right)\left(v_{F k}^{2} k^{2}+V^{2}-M\left(-V+\chi_{\pm}+\Sigma_{0 0}\right)-\left(\chi_{\pm}+\Sigma_{x x}\right)^{2}+\left(\Delta+\Sigma_{0 x}\right)^{2}\right),\\
S_{y z}^{\pm}=&-\left(\Delta+\Sigma_{y z}\right)\left(v_{F k}^{2} k^{2}+V^{2}+M\left(-V+\chi_{\pm}+\Sigma_{y y}\right)-\left(\chi_{\pm}+\Sigma_{z z}\right)^{2}+\left(\Delta+\Sigma_{y z}\right)^{2}\right).\nonumber
\end{aligned}
\end{equation}

%\onecolumngrid

\end{document}